\documentclass[12pt]{article}
\usepackage{geometry,amsmath,amssymb,enumerate,epsfig,graphicx}
\newcommand{\Omicron}{\mathrm{O}}
\newcommand{\assign}{:=}
\newcommand{\mathe}{\mathrm{e}}
\newcommand{\tmem}[1]{{\em #1\/}}
\newcommand{\tmop}[1]{\ensuremath{\operatorname{#1}}}
\newcommand{\tmtextbf}[1]{{\bfseries{#1}}}
\newcommand{\tmtextit}[1]{{\itshape{#1}}}
\newcommand{\tmtexttt}[1]{{\ttfamily{#1}}}

\begin{document}

\title{Triviality of $\phi^4_4$ theory:\\
small volume expansion and new data}
\author{
Peter Weisz\\
Max-Planck-Institut f\"ur Physik\\ 
F\"ohringer Ring 6 \\ 
80805 M\"unchen, Germany
\and
Ulli Wolff\thanks{
e-mail: uwolff@physik.hu-berlin.de} \\
Institut f\"ur Physik, Humboldt Universit\"at\\ 
Newtonstr. 15 \\ 
12489 Berlin, Germany
}
\date{}
\maketitle

\begin{abstract}
  We study a renormalized coupling $g$ and mass $m$ in four dimensional
  $\phi^4$ theory on tori with finite size $z = m L$. Precise numerical values
  close to the continuum limit are reported for $z = 1, 2, 4$, based on Monte
  Carlo simulations performed in the equivalent all-order strong coupling
  reformulation. Ordinary renormalized perturbation theory is found to work
  marginally at $z = 2$ and and to fail at $z = 1$. By exactly integrating
  over the constant field mode we set up a renormalized expansion in $z$ and
  compute three nontrivial orders. These results reasonably agree with the
  numerical data at small $z$. In the new expansion, the universal continuum
  limit exists as expected from multiplicative renormalizability. The
  triviality scenario is corroborated with significant precision.
\end{abstract}
\begin{flushright} HU-EP-10 \end{flushright}
\begin{flushright} MPP-2010-160 \end{flushright}
\begin{flushright} SFB/CCP-10-127 \end{flushright}
\thispagestyle{empty}
\newpage

\section{Introduction}

The standard view {\cite{Brezin:1976bp}}, {\cite{Luscher:1987ay}} of the
primary textbook example of a scalar field theory with quartic self-coupling
in four space-time dimensions is that it is a trivial theory. This means that
a true continuum limit of the regularized theory inevitably leads to a
non-interacting Gaussian theory. More precisely, the perturbative
renormalization group predicts that, once one is close to this limit, the
renormalized coupling asymptotically vanishes logarithmically with the cutoff.
This weak dependence keeps $\phi^4$ theory useful as a physical theory,
because we can simultaneously have only small cutoff effects but still have
sizeable interaction in an effective theory valid over a limited but large
range of length scales. In the absence of rigorous proofs of triviality in
four dimensions, it remains to follow the strategy of {\cite{Luscher:1987ay}}
and to verify in a nonperturbative lattice calculation that the perturbative
scenario actually applies close to the continuum limit, and thus to exclude
the logical possibility that it is simply irrelevant. We re-address this old
subject because numerical progress {\cite{Wolff:2009ke}} has made it
possible to render such tests more stringent with only moderate resources.
Although the situation is complicated by the coupling to other fields,
triviality is also the standard scenario for the Higgs field in the Standard
Model. The subtle nature of logarithmic effects makes a numerical check of the
triviality picture rather non-trivial in spite of the simple field structure
of $\phi^4$ theory. A number of Monte Carlo simulations have been conducted of
which {\cite{Montvay:1988uh}} is one example where the then new cluster
algorithm was used in the symmetric phase. These as well as our new
simulations took place in the Ising limit of $\phi^4$ theory, which amounts to
infinitely strong bare coupling, where we expect to find the strongest
possible interaction.

In {\cite{Wolff:2009ke}} a novel approach to simulate this model was presented
which is based on the simulation of Aizenman's random current representation
{\cite{Aizenman:1981du}}. Its numerical efficiency is closely related to the
fact that this equivalent reformulation enabled Aizenman to derive bounds that
prove triviality in $D > 4$ dimensions. For both applications it is essential
that the connected four point correlation that enters into the renormalized
coupling, can be calculated {\tmem{without}} the need of performing
cancellations and with the Lebowitz inequality {\cite{Lebowitz:1974}} being
manifest. The elimination of the otherwise unavoidable loss of significance in
the Monte Carlo allows to rather easily compute the renormalized coupling with
per mille precision close to the continuum limit. An additional decisive bonus
is the practically complete absence of critical slowing down.

The triviality conjecture in $D = 4$ is investigated here in a finite volume
renormalization scheme where the continuum limit is approached on a volume
that remains finite in units of the physical renormalized length scale. This
has the same advantage here as in asymptotically free theories
{\cite{Luscher:1991wu}}, {\cite{Luscher:1992an}}, namely that the mastering of
a large ratio of physical scales is avoided and one can thus probe the
universal continuum limit much more closely .

In this paper we employ the numerical technique described in
{\cite{Wolff:2009ke}} without changes and therefore here leave out all details
on this. The data with $z = 2$, already published in {\cite{Wolff:2009ke}},
are extended by including a lattice of size $64^4$ and by simulating two more
series of lattices at $z = 1$ and $z = 4$. In addition we report on a
computation in lattice perturbation theory that we employ to analyze the data.
This calculation goes beyond {\cite{Luscher:1987ay}} by allowing for finite
$z$ values and by computing the lattice artifacts of the Callan Symanzik
$\beta$-function up to two loops. By combining with {\cite{Luscher:1987ay}} we
also obtain the three loop term for various finite $z$. By comparing with our
data we will conclude that perturbation theory works well for large $z$ and
fails for small $z$.

A plausible reason for the failure of perturbation theory on a small torus is
that the constant zero momentum mode may receive too little Gaussian damping
to justify its perturbative treatment {\cite{Brezin:1985xx}}. We therefore
embark on an alternative approximation scheme where we treat this one mode
exactly while maintaining the perturbative expansion for all others. It turns
out that this type of expansion rearranges itself under renormalization to an
expansion in powers of $z^2$ for arbitrary values of $g / z^4$. The new
expansion is found to be applicable and accurate at small $z$.

We set up our renormalization scheme in sect.~2. Our perturbative calculation
is described in sect.~3 and app.~A and numerical results are discussed in
sect.~4. The new small volume expansion is outlined in sect.~5 and app.~B and
we conclude in sect.~6.

\section{Finite size renormalization scheme}

We define $\phi^4$ theory on a four dimensional periodic lattice of extent $L$
in all directions by the standard Euclidean action
\begin{equation}
  S = a^4 \sum_x \left\{ \frac{1}{2} \sum_{\mu} (\partial_{\mu} \phi)^2 +
  \frac{1}{2} m_0^2 \phi^2 + \frac{g_0}{4!} \phi^4 \right\} . \label{SFT}
\end{equation}
Here $\partial_{\mu} \phi (x) = [\phi (x + a \hat{\mu}) - \phi (x)] / a$ is
the standard nearest neighbor forward derivative and $\hat{\mu}$ is a unit
vector in the $\mu$ direction. From here on we mostly use lattice units and
leave out powers of $a = 1$ except for some formulas where we find that
clarity is gained by restoring explicit factors of $a$. A completely
equivalent form of the above action on the lattice is given by
\begin{equation}
  S = - 2 \kappa \sum_{x, \mu} s (x) s (x + \hat{\mu}) + \sum_x [s (x)^2 +
  \lambda (s (x)^2 - 1)^2] . \label{SLat}
\end{equation}
The relation between the two parameterizations is given by
\begin{equation}
  g_0 = 6 \lambda / \kappa^2, \hspace{1em} m_0^2 = (1 - 2 \lambda) / \kappa -
  8, \hspace{1em} \phi = \sqrt{2 \kappa} s. \label{mkappa}
\end{equation}
In the latter form it becomes manifest that for $\lambda \rightarrow \infty$
the integrations over the spin field $s (x)$ reduce to Ising sums over $s (x)
= \pm 1$.

It has become standard {\cite{Luscher:1987ay}} to take the continuum limit
along vertical lines in the $(\kappa, \lambda)$ plane by sending $\kappa$ to
its critical value $\kappa_c (\lambda)$ at fixed $\lambda$. If for the
infinite volume this limit is taken from below the theory is in the symmetric
phase.
We adhere to this
`coordinate choice' although alternative procedures are conceivable, of
course, without however changing the set of continuum theories that can be
reached.

It is now a completely well-defined procedure to perturbatively compute,
starting from (\ref{SFT}) and within the regularized theory, the (bare)
effective action{\footnote{The sign of $\Gamma$ is not uniform in the
literature. We follow the convention {\cite{Luscher:1987ay}} which leads to
$\Gamma^{(2)} (0, 0) = - m_0^2 + \Omicron (g_0)$.}} as a sum over connected
one particle irreducible graphs with propagators for external lines canceled.
At first we are interested in the 2-point vertex function $\Gamma^{(2)} (p, -
p)$. It is related to the susceptibility measured in simulations by
\begin{equation}
  \chi_2 = \sum_x \langle s (0) s (x) \rangle = \left[ - 2 \kappa \Gamma^{(2)}
  (0, 0) \right]^{- 1} .
\end{equation}
For our definition of a renormalized mass $m$ a non-zero momentum is required
for which we take the minimal one
\begin{equation}
  p_{\ast} = (2 \pi / L, 0, 0, 0) .
\end{equation}
Then the `second moment' definition of $m$ follows from the universal ratio
\begin{equation}
  \frac{\Gamma^{(2)} (0, 0)}{\Gamma^{(2)} (p_{\ast}, - p_{\ast})} =
  \frac{1}{\chi_2} \sum_x \mathe^{- i p_{\ast} x} \langle s (0) s (x) \rangle
  \assign \left( 1 + \frac{\hat{p}_{\ast}^2}{m^2} \right)^{- 1} \label{mdef}
\end{equation}
with $\hat{p}_{\mu} = 2 \sin (p_{\mu} / 2)$. In the following we define the
finite size continuum limit by holding constant the combination
\begin{equation}
  z = m L. \label{zdef}
\end{equation}
Thus for some choice of $z$ (and $\lambda$) we consider sequences of lattices
with growing $L \equiv L / a \rightarrow \infty$ \ where for each $L$ we
adjust $\kappa$ such that $z$ has the desired value. We may also say that in
this way we have defined a family of renormalization schemes, one for each
value of $z$.

The renormalized coupling is now an output observable in this procedure which
is given by another universal ratio
\begin{equation}
  g = - \frac{\chi_4}{\chi_2^2} m^4 \label{gdef}
\end{equation}
with the connected four point susceptibility
\begin{equation}
  \chi_4 = \sum_{x, y, z} \langle s (0) s (x) s (y) s (z) \rangle - 3 L^4
  \chi_2^2 = \frac{1}{(2 \kappa)^2}  \frac{\Gamma^{(4)} (0, 0, 0,
  0)}{[\Gamma^{(2)} (0, 0)]^4}
\end{equation}
or, equivalently, the 4-point vertex function $\Gamma^{(4)}$. It vanishes for
a Gaussian theory and is hence a key quantity in connection with the
triviality conjecture. An important property of our coupling is that it is
rigorously bounded in the range
\begin{equation}
  0 \leqslant g / z^4 \leqslant 2.
\end{equation}
This bound, based on {\cite{Aizenman:1981du}}, is manifestly visible in eq.
(20) of {\cite{Wolff:2009ke}} where an observable with values in  $\{0, 1\}$
is averaged with a positive weight to yield $g / (2 z^4)$.

The renormalization scheme in {\cite{Luscher:1987ay}} is defined in an
infinite volume and uses the `zero momentum' definition for the renormalized
mass $m \equiv m_R$. We make contact with this scheme by taking within our
family of schemes the simultaneous limit $z \rightarrow \infty, L \rightarrow
\infty$ at fixed $m_R = z / L$. In this limit $\hat{p}_{\ast}^2 / m^2 \simeq
(2 \pi / z)^2$ becomes arbitrarily small and (\ref{mdef}) goes over into
(2.12) of {\cite{Luscher:1987ay}}. Note that this is not a continuum limit as
long as $m_R \equiv a m_R$ is finite, in contrast to $L \equiv L / a
\rightarrow \infty$ at fixed $z$.

We are now in a position to define the Callan Symanzik $\beta$-function in our
scheme by
\begin{equation}
  \beta_z (g, a / L) = - L \frac{\partial}{\partial L} g |_{\lambda, z} .
  \label{betadef}
\end{equation}
As $L$ is integer, the derivative must be approximated by a sufficiently
accurate finite difference formula, see sect. \ref{secbarte} for further
details.

\begin{figure}[htb]
\begin{center}
  \resizebox{0.7\textwidth}{!}{\includegraphics{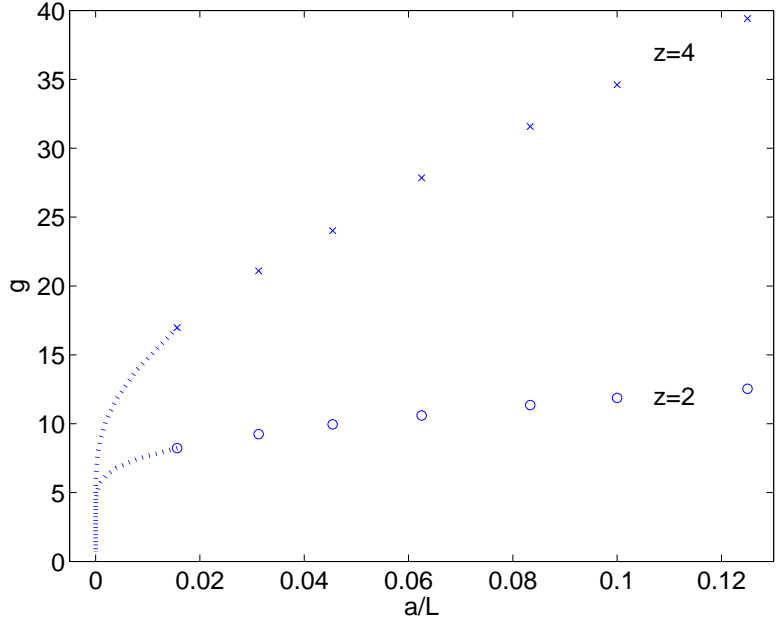}}
  \caption{Upper boundaries of the domains of renormalized coupling versus
  cutoff at $z = 2, 4$. Symbols are Monte Carlo results and 1-loop
  perturbation theory furnishes the dotted extrapolations.\label{fdomain}}
\end{center}
\end{figure}

There is a family of `curves' in the $(g, a / L)$ plane, one for each value
$\lambda \in [0, \infty)$, \ on which $\beta_z$ and other observables become
defined. We write `curve' because with only integer $L / a$ it is actually a
sequence of discrete points. If the triviality conjecture holds, then all
these curves end in the point (0,0). In figure \ref{fdomain} we see a sketch
of the expected shape of the domains for $z = 4$ and $z = 2$ based on our
nonperturbative data together with continuum extrapolations to be discussed in
section \ref{datasec}. The respective domains are the areas under the curves
if we assume that smaller $\lambda$ yield smaller $g$. An analogous curve for
$z = 1$ would be roughly around $g \sim 1.5$ before diving down to zero.

At tree level of perturbation theory we trivially find $g = g_0$ and $m =
m_0$. In terms of the natural variables in (\ref{betadef}) this reads however
\begin{equation}
  g = \frac{6 \lambda}{(1 - 2 \lambda)^2} (8 + z^2 / L^2)^2
\end{equation}
which leads to
\begin{equation}
  \beta_z = \frac{4 z^2 / L^2}{8 + z^2 / L^{^2}} g + \Omicron (g^2)
\end{equation}
as already noticed in {\cite{Luscher:1987ay}}. To avoid this tree level
lattice artefact in the $\beta$-function we change the definition of the
renormalized coupling {\cite{Wolff:2009ke}} by a term that vanishes
quadratically with the cutoff into
\begin{equation}
  \tilde{g} = g r^2 (z / L)
\end{equation}
with
\begin{equation}
  r (m) = \frac{1}{1 + m^2 / 8} .
\end{equation}
The $\beta$-function for this renormalized coupling
\begin{equation}
  \tilde{\beta}_z ( \tilde{g}, a / L) = - L \frac{\partial}{\partial L} 
  \tilde{g} |_{\lambda, z} \label{betatildedef}
\end{equation}
vanishes at tree level.

\section{Lattice perturbation theory up to two loop order\label{secLPT}}

\subsection{Artifacts of the one and two loop beta function\label{secbarte}}

In this section we discuss the asymptotic expansion of $\tilde{\beta}_z$ in
powers of $\tilde{g}$
\begin{equation}
  \tilde{\beta}_z ( \tilde{g}, a / L) = \sum_{l \ge 1}
  \tilde{b}^{(l)}_z (a / L) \tilde{g}^{l + 1} \label{betatseries}
\end{equation}
or the completely analogous formula without the tildes [which must include $l
= 0$ however]. The (perturbative) renormalizability implies the finiteness of all limits
$\lim_{a / L \rightarrow 0} \tilde{b}^{(l)}_z (a / L)$. The one and
two loop terms are scheme independent, which here means independent of $z$,
and have the universal values
\begin{equation}
  \lim_{a / L \rightarrow 0} \tilde{b}^{(1)}_z (a / L) = \bar{b}^{(1)} =
  \frac{3}{16 \pi^2}, \hspace{1em} \lim_{a / L \rightarrow 0}
  \tilde{b}^{(2)}_z (a / L) = \bar{b}^{(2)} = - \frac{17 / 3}{(16
  \pi^2)^2} . \label{unib12}
\end{equation}
In {\cite{Luscher:1987ay}} the three loop result
\begin{equation}
  \tilde{b}^{(3)}_{\infty} (0) = \bar{b}^{(3)}_{\infty} =
  \frac{26.908403}{(16 \pi^2)^3} \label{b3LW}
\end{equation}
is given for the infinite volume scheme.

In appendix \ref{appPT} we derive the coefficients of the following expansion
\begin{equation}
  \tilde{g} = \tilde{g}_0 + \tilde{p}_1 (z, L) \tilde{g}_0^2 + \tilde{p}_2 (z,
  L) \tilde{g}_0^3 + \Omicron ( \tilde{g}_0^4) \label{gtildeex}
\end{equation}
with
\begin{equation}
  \tilde{g}_0 = g_0 r^2 (m_0) .
\end{equation}
We have performed the necessary Feynman diagram sums up to $L = 100$. The only
technicality that is perhaps worth mentioning here is that by a judicious use
of both momentum and position space propagators and the fast Fourier transform
we compute all two loop diagrams by performing no more than O($L^4 \ln L)$
operations at each $L$. More details are given in app. \ref{appPT}.

\begin{figure}[htb]
\begin{center}
  \resizebox{0.45\textwidth}{!}{\includegraphics{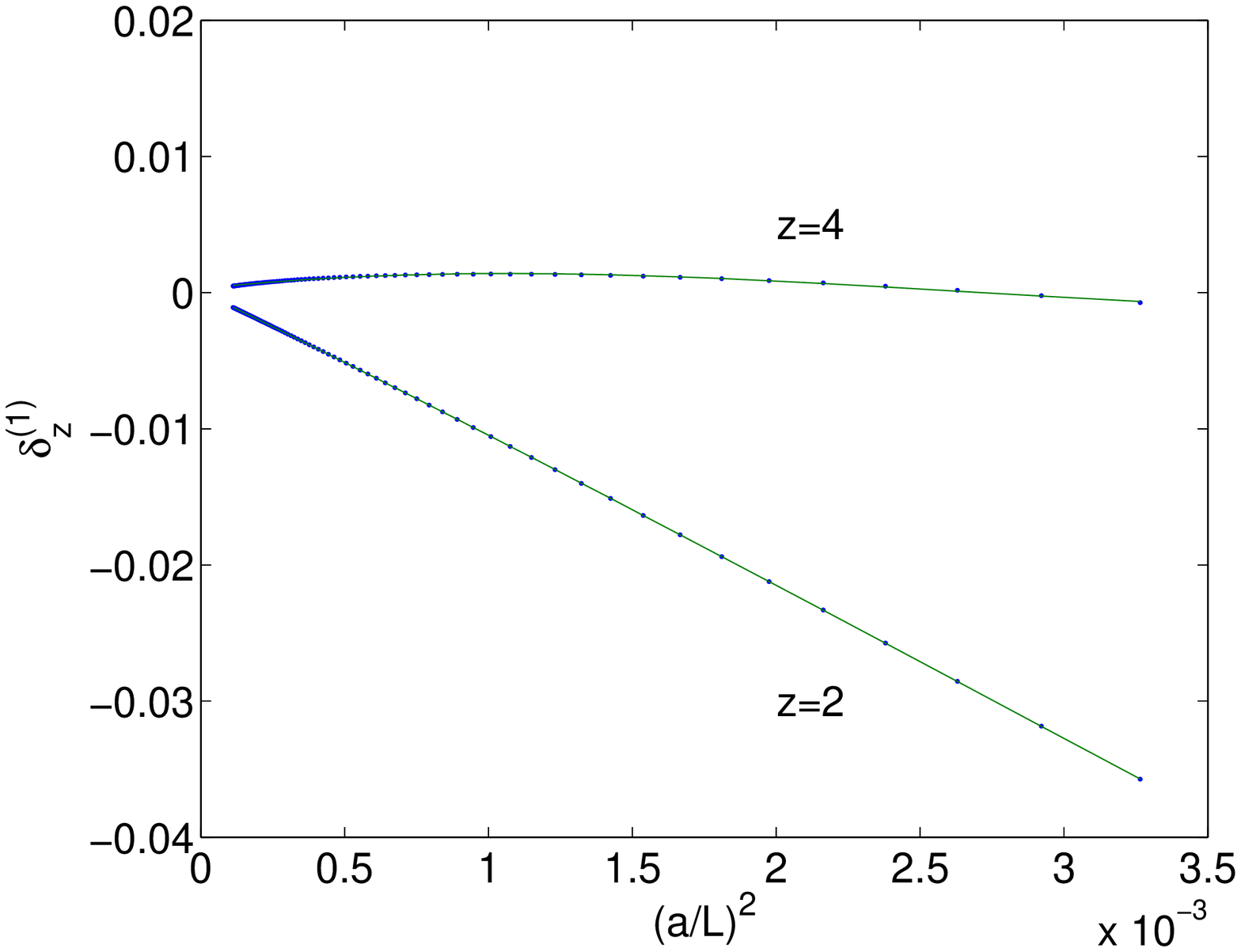}}
  \resizebox{0.45\textwidth}{!}{\includegraphics{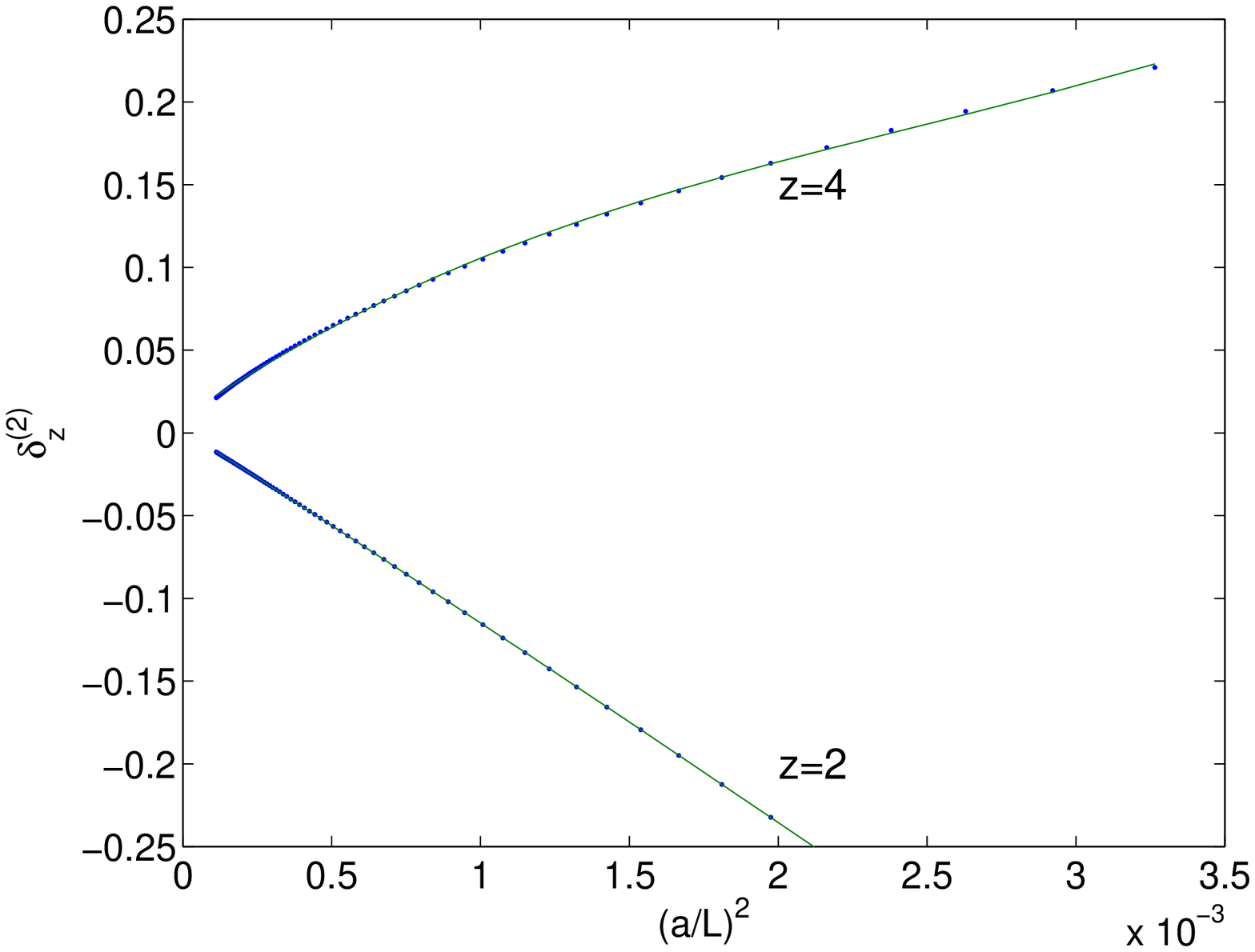}}
  \caption{Deviation from the universal values of the 1 and 2-loop
  $\beta$-function at finite lattice spacing.\label{barte}}
\end{center}
\end{figure}

Because of the relation $\tilde{g}_0 = 384 \lambda / (1 - 2 \lambda)^2$ we may
keep $\tilde{g}_0$ constant in (\ref{betatildedef}) instead of $\lambda$. Then
we obtain
\begin{equation}
  \tilde{b}^{(1)}_z (L^{- 1}) = - L \frac{\partial}{\partial L}  \tilde{p}_1
  (z, L), \hspace{1em} \tilde{b}^{(2)}_z (L^{- 1}) = - L
  \frac{\partial}{\partial L} [ \tilde{p}_2 (z, L) - \tilde{p}_1 (z, L)^2] .
\end{equation}
For several $z$ we have checked that in the Symanzik expansion in terms of
$\ln^l L L^{- 2 n}$ the exact values $\bar{b}^{(1, 2)}$ of (\ref{unib12})
emerge as leading terms ($l = 1, n = 0)$ with significant precision. With this
verified we cancel them and form the deviations
\begin{equation}
  \delta_z^{(1)} (L) = - \frac{1}{\bar{b}^{(1)}} L
  \frac{\partial}{\partial L} \left[ \tilde{p}_1 (z, L) + \bar{b}^{(1)}
  \ln L \right],
\end{equation}
\begin{equation}
  \delta_z^{(2)} (L) = - \frac{1}{\bar{b}^{(2)}} L
  \frac{\partial}{\partial L} \left[ \tilde{p}_2 (z, L) - \tilde{p}_1 (z, L)^2
  + \bar{b}^{(2)} \ln L \right] .
\end{equation}
We use the four point formula for the derivative which contributes errors of
order $L^{- 6}$ to $\delta_z^{(i)}$ which itself is expected of size O($L^{-
2}$). Results are shown in Fig. \ref{barte}. The dots are the corresponding
lattice sums while the lines are fitted third degree polynomials in $L^{- 2}$
to interpolate and represent the data.

\subsection{Three loop beta function}

We now want to relate the renormalized couplings $g$ and $g'$ referring to
schemes defined with two values $z$ and $z'$ that we freeze for a while. To
appreciate the relation we first give an `operational' description. We first
choose a value $L \equiv L / a \gg 1$. Then there are (assumed to be) bare
parameters $\lambda, \kappa$ in the scaling region that lead to renormalized
values $z = m L$ and $g$. Now {\tmem{with the bare parameters unchanged}} we
change $L \rightarrow L' \gg 1$ until $z'$ is found. In this way $g'$ and $L'
/ L$ become functions of $g$ up to scaling violations of order $L^{- 2}$ that
we neglect in this subsection. With these relations known the $\beta$ function
transforms as
\begin{equation}
  \beta_z (g) = \frac{\partial g}{\partial g'} \beta_{z'} (g') .
\end{equation}
In the end we shall be interested in the limit $z' \rightarrow \infty$ with
$g' \rightarrow g_R$ to make contact with the infinite volume scheme for which
we know the 3 loop $\beta$ function (\ref{b3LW}).

The relations just described are now analyzed in perturbation theory. Then
there is an expansion
\begin{equation}
  g = g' + P_1 (z, z') g'^2 + P_2 (z, z') g'^3 + \Omicron (g'^4) .
\end{equation}
From this form it follows that the first two expansion coefficients of the
$\beta$-functions are independent of $z$ while at three loops the connection
is
\begin{equation}
  \bar{b}_z^{(3)} - \bar{b}_{z'}^{(3)} = \bar{b}^{(1)} (P_2 -
  P_1^2) - P_1 \bar{b}^{(2)} .
\end{equation}

Using (\ref{Dm2exp}) to leading order we first derive
\begin{equation}
  L' / L = z' / z [1 + Q_1 (z, z') g' + \Omicron (g'^2)] \label{LLprime}
\end{equation}
with
\begin{equation}
  Q_1 (z, z') = \frac{1}{2}  \frac{L^2}{z^2} [q_1 (z', L z' / z) - q_1 (z, L)]
  + \Omicron (L^{- 2}) .
\end{equation}
Then we find
\begin{equation}
  P_1 (z, z') = p_1 - p_1' + \Omicron (L^{- 2}),
\end{equation}
\begin{equation}
  \hspace{1em} P_2 = p_2 - p_2' - 2 p_1' (p_1 - p_1') + \bar{b}^{(1)} Q_1
  + \Omicron (L^{- 2}) .
\end{equation}
Here $p_i, p_i'$ denote $p_i (z, L)$ and $p_i (z', L z' / z)$ respectively.
Arguments $L'$ have been eliminated with (\ref{LLprime}) and use was made of
$p_1 (z', L') = - \bar{b}^{(1)} \ln L' + \Omicron (L^0)$.

We have defined now a number of expansion coefficients that emerge as finite
continuum limits as $L \rightarrow \infty$. The Symanzik expansions of the
corresponding combinations of lattice Feynman diagrams have been analyzed by
the method given in appendix D of {\cite{Bode:1999sm}}. Results are collected
in Tab.~\ref{tab1}.
  \begin{table}[htb]
  \begin{center}
    \begin{tabular}{|l|r|r|r|r|}
      \hline
      $z, z'$ & $Q_1 \times 10^3$\phantom{0} & $P_1 \times 10^3$\phantom{0}& $P_2 \times 10^3$\phantom{0} &
      ($\bar{b}_z^{(3)} - \bar{b}_{z'}^{(3)}) \times (4 \pi)^6$\\
      \hline
      1, 2 & 210.06698\phantom{00} & $-1394.9977\phantom{00}$ & 3437.304\phantom{00} & 110315.00\phantom{00}\\
      \hline
      2, 4 & 9.7211061 & $- 80.361921$ & 12.52668 & 382.0849\\
      \hline
      4, 8 & 0.2310329 & $- 2.952908$ & 0.04461 & 0.0425\\
      \hline
      8, 16 & 0.0010522 & $- 0.024524$ & 0.00024 & $- 0.0037$\\
      \hline
      2, 3 & 8.6999022 & $- 70.199600$ & 10.75077 & 372.7864\\
      \hline
      3, 6 & 1.2399938 & $- 12.901805$ & 0.44644 & 9.4008\\
      \hline
      6, 12 & 0.0132854 & $- 0.237614$ & 0.00203 & $- 0.0651$\\
      \hline
      4, 6 & 0.2187899 & $- 2.739483$ & 0.04164 & 0.1022\\
      \hline
      8, 12 & 0.0010424 & $- 0.024189$ & 0.00022 & $- 0.0053$\\
      \hline
    \end{tabular}
    \caption{Expansion coefficients relating renormalization schemes defined
    by $z$ and $z'$ respectively. Numerical errors are beyond the digits
    quoted.\label{tab1}}
  \end{center}
  \end{table}
We see that the 3-loop coefficient $\bar{b}_z^{(3)}$ rises
very steeply for $z \lesssim 3$. The negative entries in the last column imply
that as one lowers $z$ starting from $z = 16$ (which is effectively infinite)
to smaller values, the behavior of $\bar{b}_z^{(3)}$ is not completely
monotonic. One might be tempted to think of numerical inaccuracies here, but
as far as we can tell, the negative sign seems to be significant. The table
allows to change $z$ in cycles for which we find consistency. A rough picture
is that $\bar{b}_z^{(3)}$ is constant above $z = 4$ and rises by a factor
15 for $z = 4 \rightarrow 2$ and by another factor 270 for $z = 2 \rightarrow
1$. The typical couplings will at the same time be seen below to diminish by
factors of roughly $1 / 3$ and $1 / 7$. This means that the 3-loop
contribution overall rises steeply compared to the 2-loop term with its
$z$-independent coefficient. In this way the perturbative series indicates its
breakdown for small $z$.

\section{Analysis of precise numerical data\label{datasec}}

As mentioned in the introduction, our simulations here follow in all details
those described in {\cite{Wolff:2009ke}}. In particular, for each set of
parameters we have generated a statistics of $10^6$ iterations resulting in
per mille errors. Due to the efficiency of the method this could be done in a
short time on a few up-to-date PCs.

In the next subsections we list data and plot couplings versus cutoff for $z =
4, 2, 1$. The tables contain the quantity $\mathcal{X}$ that is related to the
coupling by
\begin{equation}
  g = 2 z^4 \mathcal{X}.
\end{equation}
The evolutions of the coupling are compared with curves obtained by
integrating the renormalization group equation
\begin{equation}
  L \frac{\partial \tilde{g}}{\partial L} = - \tilde{\beta}_z ( \tilde{g}, 0)
\end{equation}
starting from the `measured' coupling at the largest $L / a = 64$. The (solid)
curves labelled with a loop order refer to the perturbative expansion of
$\tilde{\beta}_z$. We have also looked at the curves including the known one
and two loop cutoff effects, i.e. with $\tilde{\beta}_z ( \tilde{g}, 1 / L)$
on the right hand side. We find however that these small corrections do not
systematically improve the picture. Sometimes they go in the right and
sometimes in the wrong direction. We below offer a possible explanation for
this and do not include these curves in the plots. For the case $z = 2$ they
are visualized however (without $L = 64$) in Fig. 2 of {\cite{Wolff:2009ke}}.
Instead we here include additional curves (dashed) labelled LO, NLO, NNLO.
They refer to the leading, next-to-leading and next-to-next-to-leading orders
of the small volume expansion that is explained in detail in section
\ref{secPT0}.

\subsection{$z = 4$}

\begin{table}[htb]
  \begin{center}
    \begin{tabular}{|l|l|l|l|l|l|}
      \hline
      $L / a$ & $2 \kappa$ & $z$ & $\mathcal{X}$ & $\partial \mathcal{X}/
      \partial z$ & $\mathcal{X}(z = 4)$\\
      \hline
      8 & 0.141976 & 4.0025(22) & 0.07684(29) & $-$0.0556(5) & 0.07698(27)\\
      \hline
      10 & 0.144491 & 3.9956(21) & 0.06783(27) & $-$0.0518(5) & 0.06760(25)\\
      \hline
      12 & 0.145933 & 3.9991(20) & 0.06173(26) & $-$0.0489(5) & 0.06168(24)\\
      \hline
      16 & 0.147481 & 4.0045(19) & 0.05419(24) & $-$0.0441(5) & 0.05439(23)\\
      \hline
      22 & 0.148454 & 4.0004(18) & 0.04690(22) & $-$0.0387(5) & 0.04691(21)\\
      \hline
      32 & 0.1490781 & 4.0002(17) & 0.04120(21) & $-$0.0360(5) & 0.04121(20)\\
      \hline
      64 & 0.1495244 & 4.0009(15) & 0.03314(18) & $-$0.0292(7) & 0.03316(18)\\
      \hline
    \end{tabular}
  \end{center}
  \caption{Monte Carlo results for $z = 4$.\label{tabz4}}
\end{table}

\begin{figure}[htb]
\begin{center}
  \resizebox{0.9\textwidth}{!}{\includegraphics{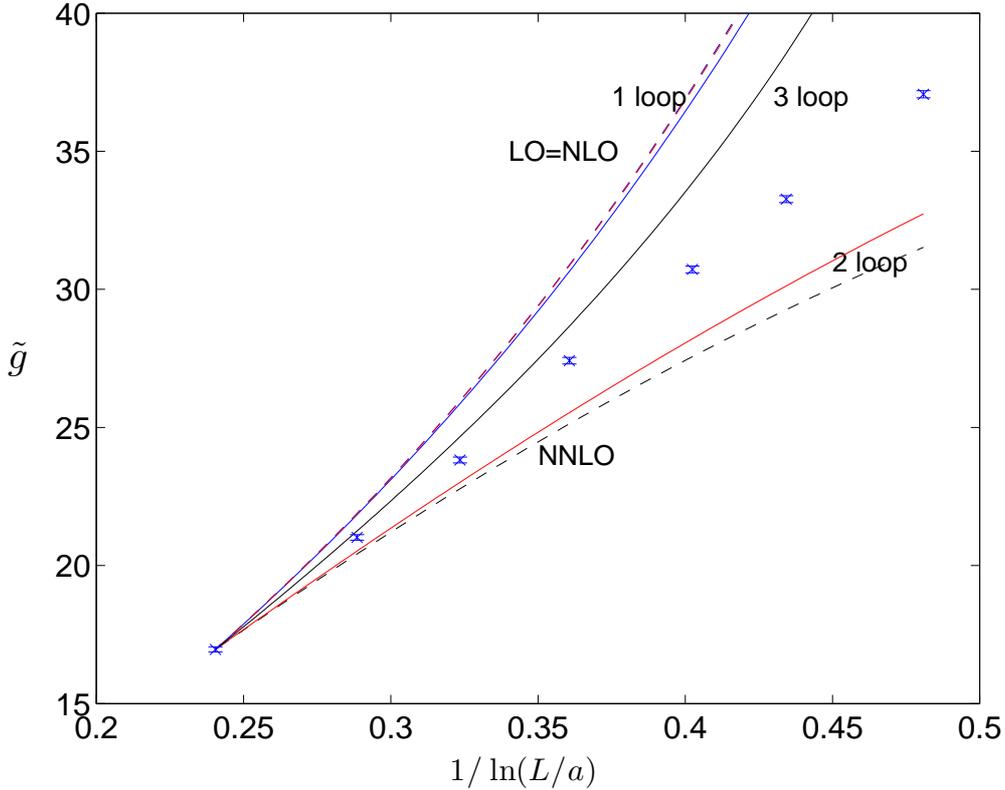}}
  \caption{Evolution of the the coupling $\tilde{g}$ with the cutoff $L / a$
  for $z = 4.$\label{rungz4}}
\end{center}
\end{figure}
Data for $z=4$ are tabulated in Tab.~\ref{tabz4} and plotted in Fig.~\ref{rungz4}.
At this $z$ value perturbation theory works as one expects: the successive
terms alternate around the nonperturbative answer and get closer to it. The
change in the coupling as the cutoff changes from $L / a = 64$ to \ $L / a =
32$ is reproduced with 1.1\% error in the 3-loop approximation (2.4\% and
3.9\% for 2- and 1-loop). In the small volume expansion LO and NLO fall almost
on top of each other and are close to 1-loop while NNLO is not far from the
2-loop result.

\subsection{$z = 2$}

\begin{table}[htb]  
  \begin{center}
    \begin{tabular}{|l|l|l|l|l|l|}
      \hline
      $L / a$ & $2 \kappa$ & $z$ & $\mathcal{X}$ & $\partial \mathcal{X}/
      \partial z$ & $\mathcal{X}(z = 2)$\\
      \hline
      8 & 0.148320 & 1.9981(27) & 0.39235(96) & $-$0.3200(14) & 0.39175(63)\\
      \hline
      10 & 0.148748 & 1.9949(26) & 0.37256(92) & $-$0.3193(14) & 0.37093(62)\\
      \hline
      12 & 0.148996 & 1.9992(26) & 0.35493(91) & $-$0.3165(15) & 0.35469(60)\\
      \hline
      16 & 0.149270 & 1.9988(25) & 0.33161(91) & $-$0.3129(16) & 0.33125(58)\\
      \hline
      22 & 0.149449 & 2.0085(24) & 0.30831(86) & $-$0.3030(16) & 0.31088(57)\\
      \hline
      32 & 0.149571 & 1.9956(24) & 0.29028(83) & $-$0.2993(20) & 0.28896(55)\\
      \hline
      64 & 0.1496564 & 1.9893(22) & 0.26016(78) & $-$0.2887(27) & 0.25706(51)\\
      \hline
    \end{tabular}  
  \end{center}
  \caption{Monte Carlo results for $z = 2$.\label{tabz2}}
\end{table}

\begin{figure}[htb]
\begin{center}
  \resizebox{0.9\textwidth}{!}{\includegraphics{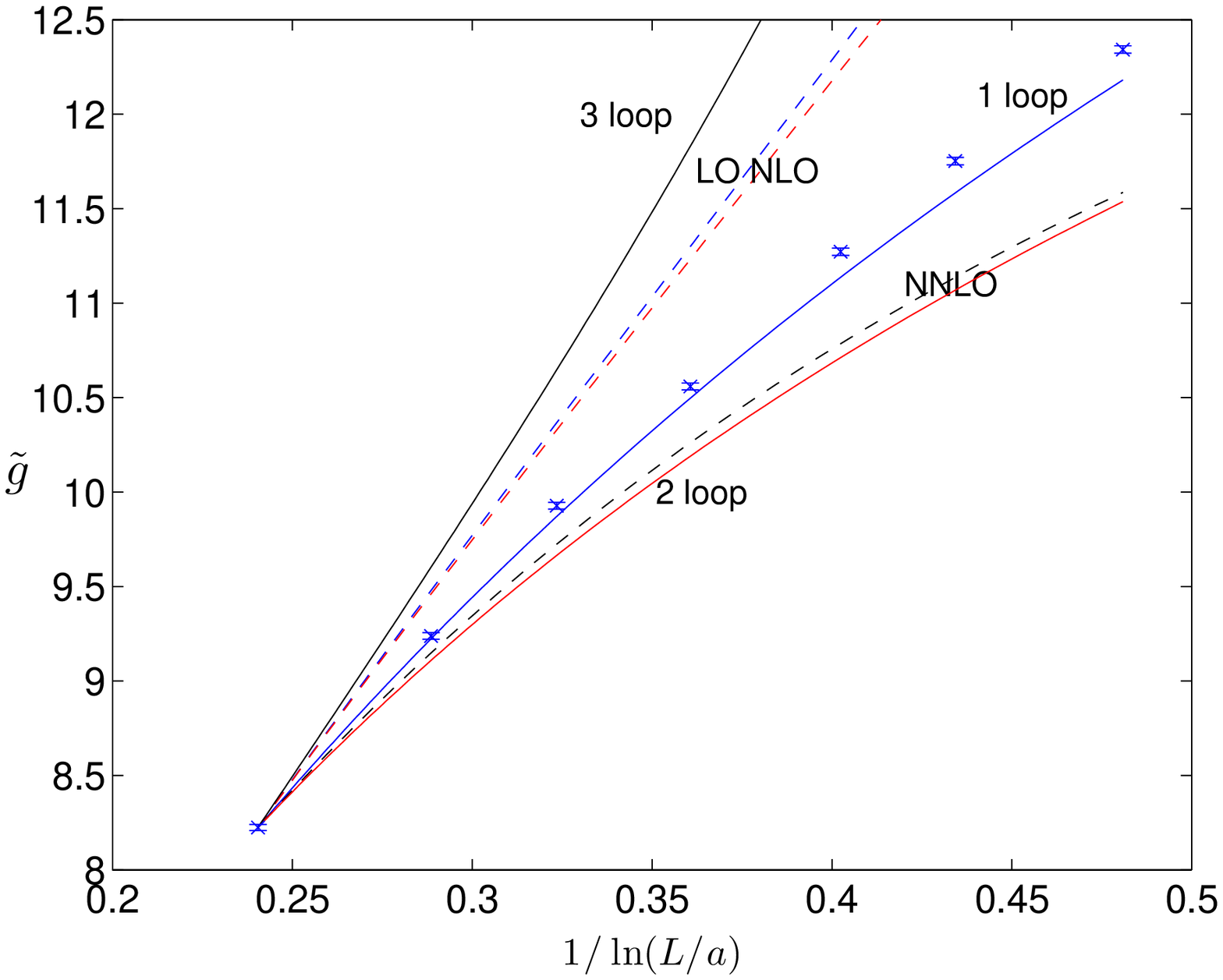}}
  \caption{Evolution of the the coupling $\tilde{g}$ with the cutoff $L / a$
  for $z = 2$.\label{rungz2}}
\end{center}
\end{figure}
Data for $z=2$ are tabulated in Tab.~\ref{tabz2} and plotted in Fig.~\ref{rungz2}.
For $z = 2$ the one loop result happens to be exact within errors (0.2\%),
while 2- and 3-loop are off by 1.4\% and 3.9\% respectively for the same
benchmark as discussed in the previous subsection. As already concluded in
{\cite{Wolff:2009ke}} this is a limiting case for perturbation theory as an
asymptotic expansion, where higher loops do not help at all any more. The
precision of the small volume expansion is 2.6\%, 2.4\%, 1.0\% \ for LO, NLO,
NNLO respectively.

\subsection{$z = 1$}

\begin{table}[htb]  
  \begin{center}
    \begin{tabular}{|l|l|l|l|l|l|}
      \hline
      $L / a$ & $2 \kappa$ & $z$ & $\mathcal{X}$ & $\partial \mathcal{X}/
      \partial z$ & $\mathcal{X}(z = 1)$\\
      \hline
      8 & 0.151670 & 1.0020(25) & 0.79953(84) & $-$0.4444(30) & 0.80040(85)\\
      \hline
      10 & 0.150900 & 0.9997(24) & 0.78776(88) & $-$0.4641(30) & 0.78764(83)\\
      \hline
      12 & 0.150498 & 0.9999(24) & 0.77930(92) & $-$0.475(30) & 0.77926(82)\\
      \hline
      16 & 0.150118 & 1.0032(22) & 0.76121(96) & $-$0.5074(31) & 0.76281(80)\\
      \hline
      22 & 0.149907 & 0.9961(22) & 0.75058(101) & $-$0.5213(32) & 0.74854(78)\\
      \hline
      32 & 0.149787 & 1.0032(21) & 0.72936(106) & $-$0.5436(50) & 0.73110(75)\\
      \hline
      64 & 0.1497143 & 0.9993(20) & 0.70179(110) & $-$0.5797(48) & 0.70141(71)\\
      \hline
    \end{tabular}
  \end{center}
  \caption{Monte Carlo results for $z = 1$.\label{tabz1}}
\end{table}

\begin{figure}[htb]
\begin{center}
  \resizebox{0.9\textwidth}{!}{\includegraphics{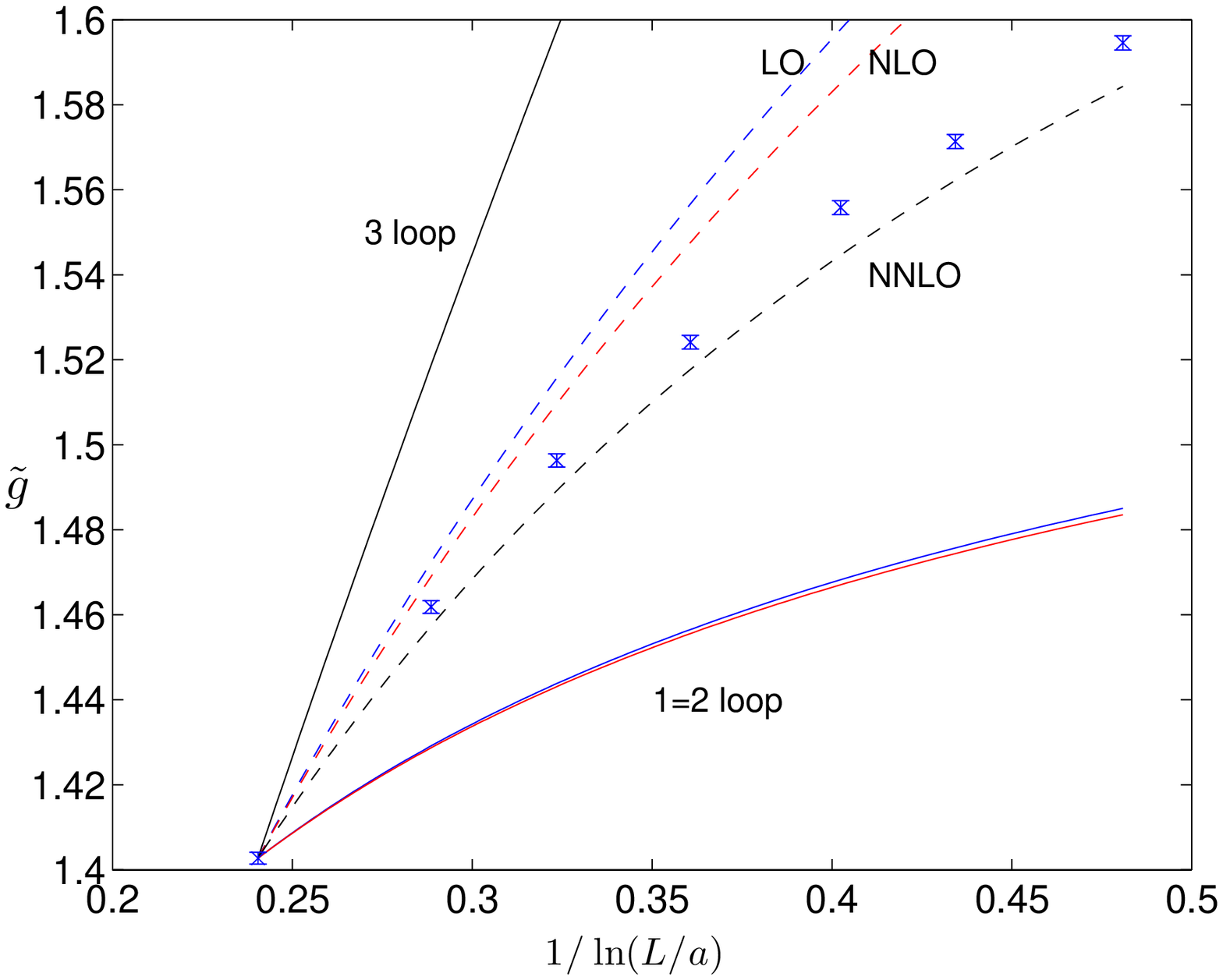}}
  \caption{Evolution of the the coupling $\tilde{g}$ with the cutoff $L / a$
  for $z = 1$.\label{rungz1}}
\end{center}
\end{figure}
Data for $z=1$ are tabulated in Tab.~\ref{tabz1} and plotted in Fig.~\ref{rungz1}.
At $z = 1$ perturbation theory is not useful anymore. The coupling itself (but
not $g / z^4$ on its natural scale) has small values and therefore the 2-loop
contribution, whose coefficient cannot depend on $z$, is tiny. The very large
3-loop term knocks the approximation far to the other side. The small volume
expansion exhibits deviations of 0.7\%, 0.5\%, 0.3\% for LO, NLO, NNLO.

\subsection{Lattice artifacts and perturbation theory}

In this subsection we offer a short discussion whether or not we expect the
lattice artifacts to be well represented by perturbation theory. This may be
questionable even in the regime where the continuum behavior {\tmem{is}}
reproduced.

We want to analyze closer the perturbative procedure that we have applied,
which is however the conventional one. To derive the expansion coefficients
(\ref{betatseries}) we consider $g_0$ as an arbitrarily small expansion
parameter and correspondingly expand around $\phi \equiv 0$ in (\ref{SFT}).
Then we re-expand the truncated series in the renormalized coupling $g$ or
$\tilde{g}$. This series is then held against lattice data produced with
$\lambda$ or $g_0$ such that the {\tmem{bare}} lattice action has a pair of
nonzero constant minima. This is so in the most extreme way for the Ising
limit. Renormalized perturbation theory is expected to describe the universal
physics in spite of the apparent contradiction as long as successive terms in
the $g$ expansion (at small $a / L$) look reasonably `convergent'. A possible
explanation is as follows. We may simultaneously consider a lattice that is
coarser but still in the scaling region tuned to the same (matched)
renormalized parameters. On this lattice the bare $g_0$ will be smaller and
the expansion may also be {\tmem{naively}} justified. Up to differing small
scaling violations the universal physics will then agree on both lattices. One
could consider the extension of this matching to the leading cutoff effects in
the spirit of the Symanzik effective action. But then one would need further
improvement terms on the coarser lattice to reproduce the cutoff effects of
the finer one. As an alternative qualitative argument we could imagine a
block-spin coarsening of the original lattice. Then we would expect to arrive
at an effective action on the coarse lattice consisting of the relevant terms
of (\ref{SFT}) with small $g_0$ and additional contributions corresponding to
Symanzik terms. Therefore the relation between cutoff effects computed in
perturbation theory with the standard action (\ref{SFT}) as sketched above and
those of the nonperturbative results may not be so straightforward in our
opinion.

\section{Small volume perturbation theory\label{secPT0}}

In this section we introduce a modified perturbative expansion which will turn
out to lead to a systematic expansion in the finite size scaling variable $z$
defined in (\ref{zdef}).

In our perturbative calculation the Gaussian damping of the low momentum modes
on the torus is controlled by $\omega^2 \cong m_0^2 + (2 \pi / L)^2 
\vec{n}^2$ with $\vec{n}$ having small integer components. For small $z$ with
$m_0 = z / L + \Omicron (g_0)$ the $\vec{n} = 0$ mode receives little damping
and ordinary perturbation theory may be a bad starting point
{\cite{Brezin:1985xx}}. We therefore split up the lattice field
\begin{equation}
  \phi (x) = \frac{1}{L g_0^{1 / 4}}  \bar{\phi} + \eta (x), \hspace{1em}
  \sum_x \eta (x) = 0.
\end{equation}
We decompose the action (\ref{SFT}) into three parts, $S = S_{0,
\bar{\phi}} + S_{0, \eta} + S_1$,
\begin{equation}
  S_{0, \bar{\phi}} = \frac{1}{2} z_0^2 \bar{\phi}^2 + \frac{1}{4!}
  \bar{\phi}^4, \label{S0phi}
\end{equation}
\begin{equation}
  S_{0, \eta} = \frac{1}{2} \sum_{x,\mu} \left\{ (\partial_{\mu} \eta)^2 +
  g_0^{1 / 2} \frac{2 z_0^2 + \bar{\phi}^2}{2 L^2} \eta^2 \right\}
  \label{S0eta}
\end{equation}
and
\begin{equation}
  S_1 = \sum_x \left\{  \frac{g_0^{3 / 4}}{3! L}  \bar{\phi} \eta^3 +
  \frac{g_0}{4!} \eta^4 \right\} \label{S1}
\end{equation}
with
\begin{equation}
  z_0^2 = \frac{L^2 m_0^2}{\sqrt{g_0}} .
\end{equation}
For every cutoff $L / a$ we consider now $z_0^2, g_0$ as bare input and want
to compute from them $z$ and $g$. The idea is to treat $S_{0, \cdot}$ exactly
and $S_1$ as a perturbation, i.e. $g_0$ is small at finite $z_0^2$. In the end
the propagator implied by (\ref{S0eta}) will also be expanded in $g_0^{1 /
2}$. Alternatively it could be included as a two $\eta$ vertex in $S_1$ from
the beginning which we found less efficient however.

The second moment renormalized mass definition (\ref{mdef}) implies
\begin{equation}
  \frac{z^2}{1 + z^2 / K_L} = \sqrt{g_0}  \frac{\hat{p}^2_{\ast}
  \tilde{\Delta} (p_{\ast})}{\langle \bar{\phi}^2 \rangle} \label{zrel}
\end{equation}
with
\begin{equation}
  K_L = L^2 \hat{p}_{\ast}^2 = 4 \pi^2 - \frac{4 \pi^4}{3 L^2} + \Omicron
  (L^{- 4}) .
\end{equation}
and
\begin{equation}
  \tilde{\Delta} (p) = \sum_x \mathe^{- i p x} \langle \eta (0) \eta (x)
  \rangle .
\end{equation}
By doing perturbation theory in $\sqrt{g_0}$ we first produce the coefficients
of the expansions
\begin{equation}
  g / z^4 = 3 - \frac{\langle \bar{\phi}^4 \rangle}{(\langle
  \bar{\phi}^2 \rangle)^2} = \sum_{n \geqslant 0} c_n (z_0^2, L) g_0^{n /
  2} \label{ging0}
\end{equation}
and from (\ref{zrel})
\begin{equation}
  z^2 = \sqrt{g_0} \sum_{n \geqslant 0} d_n (z_0^2, L) g_0^{n / 2} .
  \label{zting0}
\end{equation}
The $\beta$ function is computed as
\begin{equation}
  \beta_z = - L \frac{\partial g}{\partial L}  |_{z, g_0} .
\end{equation}
For the derivative we use
\begin{equation}
  L \frac{\partial}{\partial L}  |_{z, g_0} = L \frac{\partial}{\partial L} -
  \rho (z_0^2, L) \frac{\partial}{\partial z_0^2} \label{ztder}
\end{equation}
where all partial derivatives on the right hand side are now taken with
respect to the set $(z_0, L, g_0)$ and
\begin{equation}
  \rho (z_0^2, L) = L \frac{\partial z^2}{\partial L} \left[  \frac{\partial
  z^2}{\partial z_0^2} \right]^{- 1}
\end{equation}
has been introduced.

If we invert (\ref{zting0}) to express $\sqrt{g_0}$ as a series in $z^2$ we
obtain
\begin{equation}
  g / z^4 = \sum_{n \geqslant 0} e_n (z_0^2, L) z^{2 n} \label{ginzt}
\end{equation}
from (\ref{ging0}) and we arrive at the intermediate form
\begin{equation}
  \beta_z = z^6 \sum_{n \geqslant 0} f_n (z_0^2, L) z^{2 n} . \label{betainzt}
\end{equation}
Below we now refer to results whose derivation is described in some detail in
appendix \ref{appMPT}.

The first coefficients $e_0, e_1$ are given by
\begin{equation}
  e_0 = 3 - \frac{\mu_4}{\mu^2} = 3 \frac{\mu^2 + 2 \mu z_0^2 - 2}{\mu^2}
\end{equation}
and
\begin{equation}
  e_1 = - \frac{3 C_1}{2 \mu^2}  \left\{ \mu^3 z_0^2 + 2 \mu^2 (3 z_0^4 -
  2) - 18 \mu z_0^2 + 12 \right\} \equiv \frac{1}{2} \mu C_1
  \frac{\partial}{\partial z_0^2} e_0 \label{e1e0}
\end{equation}
with the second moment computed with (\ref{S0phi})
\begin{equation}
  \mu (z_0^2) = \langle \bar{\phi}^2 \rangle_{0, \phi}
\end{equation}
and the constant $C_1$ defined in (\ref{Ckdef}). For given $g, z$ we solve for
$\bar{z}_0^2$ in
\begin{equation}
  g / z^4 = e_0 ( \bar{z}_0^2)
\end{equation}
which gives $\bar{z}_0$ the status of a renormalized parameter. One finds
for example $\bar{z}_0^2 \approx 0.303$ ($\mu \approx 1.25$) for $g = 10,
z = 2$. The bare quantity $z_0^2$ is then written as a power series in $z^2$
by solving (\ref{ginzt})
\begin{equation}
  z_0^2 = \bar{z}_0^2 + \sum_{n \geqslant 1} h_n ( \bar{z}_0^2, L)
  z^{2 n} . \label{z0series}
\end{equation}
The lowest order terms are
\begin{eqnarray}
  h_1 & = & - e_1 \left[ \frac{\partial e_0}{\partial z_0^2} \right]^{- 1} = -
  \frac{1}{2} \mu C_1, \label{shift} \\
  h_2 & = & - \left[ e_2 + h_1 \frac{\partial e_1}{\partial z_0^2} +
  \frac{1}{2} \frac{\partial^2 e_0}{\partial (z_0^2)^2} h_1^2 \right] \left[
  \frac{\partial e_0}{\partial z_0^2} \right]^{- 1}, \\
  h_3 & = & - \left[ e_3 + h_1 \frac{\partial e_2}{\partial z_0^2} +
  \frac{h_1^2}{2} \frac{\partial^2 e_1}{\partial (z_0^2)^2} + \frac{h_1^3}{6}
  \frac{\partial^3 e_0}{\partial (z_0^2)^3} + \frac{\partial^2 e_0}{\partial
  (z_0^2)^2} h_1 h_2 + h_2 \frac{\partial e_1}{\partial z_0^2} \right] \left[
  \frac{\partial e_0}{\partial z_0^2} \right]^{- 1} 
\end{eqnarray}
with all quantities on the right hand sides taken at $\bar{z}_0^2$. To
finally obtain the $\beta$ function in the form which is expected to possess a
continuum limit we eliminate $z_0^2$ from (\ref{betainzt}) to obtain
\begin{equation}
  \beta_z (g) = z^8 (B_1 + B_2 z^2 + B_3 z^4) + \Omicron (z^{14})
\end{equation}
where the $B_k$ are functions of $g / z^4$ (via $\bar{z}_0^2$) only after
neglecting lattice artifacts proportional to $L^{- 2}$. This also is the
reason why there is no O($z^6$) contribution, since
\begin{equation}
  f_0 = \Omicron (L^{- 2})
\end{equation}
holds. This can be understood by noting that the leading order $g_0^{1 / 2}$
correction is essentially a mass renormalization, the term proportional to
$\bar{\phi}^2$ in (\ref{lP0}). This leads to relations
\begin{equation}
  c_1 = \frac{1}{2} C_1 \frac{d c_0}{d z_0^2}, \hspace{1em} d_1 = \frac{1}{2}
  C_1 \frac{d d_0}{d z_0^2} - \frac{\mu^2 + 2 z_0^2 \mu - 2}{2 K_L \mu^2} 
\end{equation}
and also to the last equality in (\ref{e1e0}). With these identities the
cancellation that leaves only cutoff effects for $f_0$ can be shown. The first
three $B_k$ are
\begin{eqnarray}
  B_1 &=& f_1 + h_1 \frac{\partial f_0}{\partial z_0^2}, \\
  B_2 &=& f_2 + h_1 \frac{\partial f_1}{\partial z_0^2} + h_2 \frac{\partial
  f_0}{\partial z_0^2} + \frac{1}{2} h_1^2 \frac{\partial^2 f_0}{\partial
  (z_0^2)^2}, \\
  B_3 &=& f_3 + h_1 \frac{\partial f_2}{\partial z_0^2} + \frac{h_1^2}{2}
  \frac{\partial^2 f_1}{\partial (z_0^2)^2} + \frac{h_1^3}{6} \frac{\partial^3
  f_0}{\partial (z_0^2)^3} + h_1 h_2 \frac{\partial^2 f_0}{\partial (z_0^2)^2}
  + h_2 \frac{\partial f_1}{\partial z_0^2} + h_3 \frac{\partial f_0}{\partial
  z_0^2}\nonumber \\
\end{eqnarray}
where the limits $L \rightarrow \infty$ are understood on the right hand
sides. It turns out that these combinations are such, that all divergences
cancel and only finite universal results enter into the continuum limits of
the $B_k$, see app. \ref{appMPT} for more details.

\begin{figure}[htb]
\begin{center}
  \resizebox{0.8\textwidth}{!}{\includegraphics{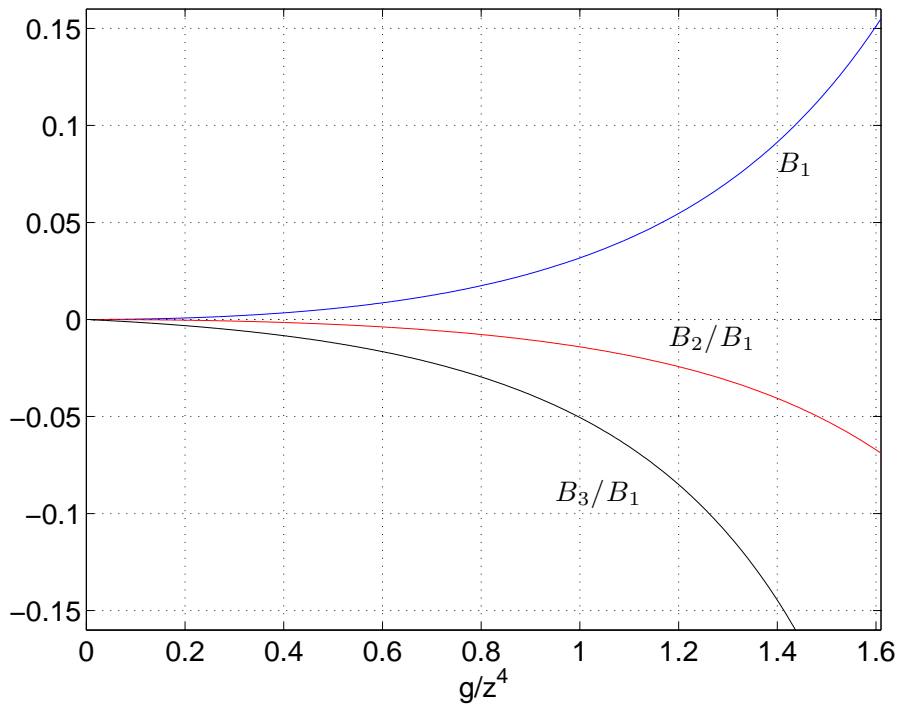}}
  \caption{Curves of $B_1$, $B_2 / B_1$ and $B_3 / B_1$ versus $g / z^4$
  produced by varying $\bar{z}_0^2$ between $10$ and $- 1.2$.
  \label{B1fig}}
\end{center}
\end{figure}

Our first result is

\begin{equation}
  B_1 = - \frac{9}{16 \pi^2} \frac{\mu^3 \bar{z}_0^2 + 2 \mu^2 (3
  \bar{z}_0^4 - 2) - 18 \mu \bar{z}_0^2 + 12}{\mu^2 + 6 \mu
  \bar{z}_0^2 - 6} . \label{B1res}
\end{equation}
For large $\bar{z}_0^2$ the fluctuations of $\bar{\phi}$ become
Gaussian and $\mu$ may be expanded as in (\ref{muexp}). We then find the
leading large $\bar{z}_0^2$ behavior
\begin{equation}
  e_0 \simeq \frac{1}{\bar{z}_0^4}, \hspace{1em} B_1 \simeq \frac{3}{16
  \pi^2}  \frac{1}{\bar{z}_0^8},
  \hspace{1em} z^8 B_1 \simeq \bar{b}_1 g^2
  \hspace{1em} ( \bar{z}_0^2 \rightarrow \infty)
\end{equation}
such that in this limit we recover the perturbative result 
with the one loop coefficient (\ref{unib12}). More
generally we may run $\bar{z}_0^2$ through some range and parametrically
generate the graph of $B_1$ versus $g / z^4$ as shown in Fig. \ref{B1fig}.

In the next order we find
\begin{equation}
  B_2 = - \frac{4}{9} B_1^2 . \label{B2res}
\end{equation}
At the moment we see this form on the basis of our series expansion for $B_2$
and cannot give a simple reason for this simple relation. In the perturbative
limit we thus obtain $B_2 z^{10} \simeq - (4 / 9) \bar{b}_1^2 g^4 / z^6$
which is a contribution that in ordinary perturbation theory could only come
at the 3-loop level.

Finally we find
\begin{equation}
  B_3 = - \frac{B_1}{16 \pi^2} \left[ \frac{1}{4 \pi^2} R_3 + \frac{17}{9}
  \mu^2 \right] \label{B3parts}
\end{equation}
with
\begin{eqnarray}
  R_3 & = &  \frac{1}{(\mu^2 + 6 \mu \bar{z}_0^2 - 6)^2} \times
  \big\{ \mu^8 + 16 \mu^7 \bar{z}_0^2 + 4 \mu^6 (20 \bar{z}_0^4
  - 1) + 4 \mu^5 \bar{z}_0^2 (24 \bar{z}_0^4 - 13) \nonumber\\
  &  & - 16 \mu^4 (9 \bar{z}_0^8 + 3 \bar{z}_0^4 + 4) + 48 \mu^3
  \bar{z}_0^2 (15 \bar{z}_0^4 - 7) - 228 \mu^2 (5 \bar{z}_0^4 -
  1) \nonumber\\
  &  & + 1296 \mu \bar{z}_0^2 - 432 \big\} .  \label{B3res}
\end{eqnarray}
Also these factorizations have been found by inspection only. The perturbative
limit
\begin{equation}
  R_3 \simeq - \frac{4}{\bar{z}_0^8} \hspace{1em} ( \bar{z}_0^2
  \rightarrow \infty)
\end{equation}
yields $B_3 z^{12} \simeq \bar{b}_2 g^3$ with
the term with $R_3$ not contributing to the leading order. 
We thus recover the 2-loop term (\ref{unib12}).

The running with the leading (LO, $B_1$), next-to-leading (NLO, $B_{1, 2}$)
and next-to-next-to-leading (NNLO, $B_{1, 2, 3}$) order $\beta$ function in
the small volume expansion is shown by the dashed lines in Figs. \ref{rungz4}
, \ref{rungz2} and \ref{rungz1}.

\section{Conclusions}

We have reported and discussed data for a reasonably defined renormalized
coupling $g$ in $\phi^4$ theory in finite size systems with $z = 1, 2, 4$ on
lattices up to $64^4$. All simulations took place at infinite bare coupling.
The dependence of $g$ on the cutoff $a / L$ at $z = 4$ was found to be well
described by perturbation theory with the first three loop orders becoming
successively more accurate. Thus there is all reason to trust the perturbative
continuation of these curves all the way to vanishing lattice spacing $a$
which implies triviality. At $z = 2$ on the other hand it was already found in
{\cite{Wolff:2009ke}} that a (presumably accidental) excellent agreement of
the nonperturbative evolution with the one loop approximation holds, but that
it only deteriorates when further terms are included. We now have added the
finding that at $z = 1$ the standard perturbative description fails
completely.

The suspicion that the constant mode on the torus is the culprit responsible
for the failure has led us to work out an expansion where this one mode is
treated nonperturbatively. Upon renormalization this has led us to an
expansion of the $\beta$ function in powers of $z$ itself for finite values of
$g / z^4$ for which we rigorously know that their domain is contained in the
interval $[0, 2]$. We construct completely explicitly the first three orders
of this expansion, neglecting only order $z^{14}$ and higher, and find it to
work very well at our smaller $z$ values. Since this approximation also yields
a positive $\beta$ function, the coupling evolution may be continued to
vanishing $a$ and triviality follows once more. The intermediate value $z = 2$
seems somewhat problematic for both expansions and one loop truncated
perturbation theory happens to be better than the $z$-expansion in this
difficult range which was unerringly chosen as the first application of the
new simulation technique in {\cite{Wolff:2009ke}}.

We end on a more general remark. Many structural properties of quantum fields
theory like renormalizability and the existence of the continuum limit are (in
favorable cases) based on proofs to all orders of perturbation theory. It is
not so often emphasized, that for nonperturbative calculations on the lattice
one has to assume that this holds true also beyond perturbation theory,
although we know that the latter can become numerically completely irrelevant,
even if one has command over high orders. In our small volume expansion which
may perhaps be seen as capturing a glimpse of nonperturbative behavior in an
analytic treatment, we could confirm these standard assumptions.

{\noindent}\tmtextbf{Acknowledgments}. We thank Jean Zinn-Justin for
discussions in the early stage of this project. U. W. has enjoyed hospitality
and support by the Max Planck (Werner Heisenberg) Institut in M\"unchen.
Financial support of the DFG via SFB transregio 9 is acknowledged.

\appendix\section{Perturbative expansion\label{appPT}}

We here report details on the computation of $\tilde{p}_{1, 2} (z, L)$
entering in section \ref{secLPT}. In terms of 1PI vertex functions our
renormalized parameters read
\begin{equation}
  Z^{- 1} = [\Gamma^{(2)} (0, 0) - \Gamma^{(2)} (p_{\ast}, - p_{\ast})] /
  \hat{p}_{\ast}^2 ,\label{Zdef}
\end{equation}
\begin{equation}
  m^2 = - Z \Gamma^{(2)} (0, 0),
\end{equation}
\begin{equation}
  g = - Z^2 \Gamma^{(4)} (0, 0, 0, 0)
\end{equation}
with the wave function renormalization factor $Z$ and all quantities at finite
$L$. Standard bare perturbation theory gives
\begin{equation}
  \Gamma^{(2)} (p, - p) = - m_0^2 - \hat{p}^2 - g_0 \frac{1}{2} J_1 + g_0^2
  \frac{1}{4} J_1 H_1 + g_0^2 \frac{1}{6} J_2 ( \text{$m_0^2$,} p) + \Omicron
  (g_0^3),
\end{equation}
\begin{equation}
  \Gamma^{(4)} (0, 0, 0, 0) = - g_0 + g_0^2  \frac{3}{2} H_1 - g_0^3 \left( 3
  H_{2 a} + \frac{3}{4} H_{2 b} + \frac{3}{2} H_{2 c} \right) + \Omicron
  (g_0^3) .
\end{equation}
The capital letters stand for the usual Feynman diagrams for the two and four
point functions up to two loops. With all external lines at zero momentum
except in $J_2 (m_0^2, p)$ they are regarded as functions of $m_0^2$ (and $L$)
at this stage. Their actual evaluation proceeds via the following sequence of
steps,
\begin{equation}
  \tilde{G} (p) = \frac{1}{\hat{p}^2 + m_0^2},
\end{equation}
\begin{equation}
  G (x) = \frac{1}{L^4} \sum_p \mathe^{i p x} \tilde{G} (p),
\end{equation}
\begin{equation}
  J_1 (m_0^2) = G (0),
\end{equation}
\begin{equation}
  H_1 (m_0^2) = \frac{1}{L^4} \sum_p [ \tilde{G} (p)]^2 = \sum_x [G (x)]^2
  \hspace{1em} (\tmop{Plancherel}) .
\end{equation}
The momentum sums run over all values $p_{\mu} = 2 \pi n_{\mu}/ L$,
$n_{\mu} = 0, 1, 2, \ldots, L - 1$. Next we compute
\begin{equation}
  \widetilde{G^n} (p) = \sum_x [G (x)]^n \mathe^{- i p x}, \hspace{1em} n = 2,
  3, \label{powtilde}
\end{equation}
and use it in
\begin{equation}
  H_{2 a} (m_0^2) = \frac{1}{L^4} \sum_p \widetilde{G^2} (p) [ \tilde{G}
  (p)]^2,
\end{equation}
\begin{equation}
  H_{2 b} (m_0^2) = [ \widetilde{G^2} (0)]^2 = H_1 (m_0^2)^2,
\end{equation}
\begin{equation}
  H_{2 c} (m_0^2) = J_1 \frac{1}{L^4} \sum_p [ \tilde{G} (p)]^3,
\end{equation}
\begin{equation}
  J_2 (m_0^2, p) = \widetilde{G^3} (p) .
\end{equation}
The Fourier transformations are performed as FFT on one coordinate direction
after another, schematically like
\begin{equation}
  \tilde{G} (p) \equiv F (p_0, p_1, p_2, p_3) \rightarrow F' (x_0, p_1, p_2,
  p_3) \rightarrow F'' (x_0, x_1, p_2, p_3) \rightarrow \ldots . \rightarrow G
  (x)
\end{equation}
at a total computational complexity of $D L^4 \ln L$ only.

With these expressions we can write (omitting the remainders$\ldots . +
\Omicron (g_0^3)$)
\begin{equation}
  Z = 1 + \frac{g_0^2}{6 \hat{p}_{\ast}^2} [J_2 (m_0^2, p_{\ast}) - J_2
  (m_0^2, 0)] .
\end{equation}
and
\begin{equation}
  \Delta m^2 = m_0^2 - m^2 = - \frac{g_0}{2} J_1 + \frac{g_0^2}{4} J_1 H_1 +
  \frac{g_0^2}{6} \left[ (1 + m_0^2 / \hat{p}_{\ast}^2) J_2 - (m_0^2 /
  \hat{p}_{\ast}^2) J_2 (m_0^2, p_{\ast}) \right]
\end{equation}
\begin{equation}
  g = g_0 - g_0^2  \frac{3}{2} H_1 + g_0^3 \left( 3 H_{2 a} + \frac{3}{4} H_{2
  b} + \frac{3}{2} H_{2 c} + \frac{1}{3 \hat{p}_{\ast}^2} [J_2 (m_0^2,
  p_{\ast}) - J_2] \right) .
\end{equation}
In these expressions the mass is still $m_0$ in all diagrams. In order to
obtain $g$ as a function of $g_0$ and $m^2$ we have to combine the last two
lines to eliminate $m_0^2$ on the right hand sides. To the order considered
and using
\begin{equation}
  \frac{d J_1}{d m_0^2} = - H_1, \hspace{1em} J_1 \frac{d H_1}{d m_0^2} = - 2
  H_{2 c} \label{H1der}
\end{equation}
we arrive at
\begin{equation}
  \Delta m^2 = q_1 (z, L) g_0 + q_2 (z, L) g_0^2 \label{Dm2exp}
\end{equation}
with
\begin{eqnarray}
  q_1 (z, L) & = & - \frac{1}{2} J_1 (m^2) \\
  q_2 (z, L) & = & \frac{1}{6}  \left[ (1 + m^2 / \hat{p}_{\ast}^2) J_2 (m^2,
  0) - (m^2 / \hat{p}_{\ast}^2) J_2 (m^2, p_{\ast}) \right],
\end{eqnarray}
and then
\begin{equation}
  g = g_0 + p_1 (z, L) g_0^2 + p_2 (z, L) g_0^3
\end{equation}
with
\begin{eqnarray}
  p_1 (z, L) & = & - \frac{3}{2} H_1 (m^2), \\
  p_2 (z, L) & = & 3 H_{2 a} (m^2) + \frac{3}{4} H_{2 b} (m^2) + \frac{1}{3
  \hat{p}_{\ast}^2} [J_2 (m^2, p_{\ast}) - J_2 (m^2, 0)] . 
\end{eqnarray}

We now finally change from the expansion of $g$ in powers of $g_0$ to the one
of $\tilde{g}$ in $\tilde{g}_0$ of (\ref{gtildeex}) with the corresponding
coefficients. Some straightforward steps lead to
\begin{equation}
  \tilde{q}_1 = r^{- 2} q_1, \hspace{1em} \tilde{q}_2 = r^{- 4} q_2 +
  \frac{r^{- 3}}{4} q_1^2
\end{equation}
and
\begin{equation}
  \tilde{p}_1 = r^{- 2} p_1 + \frac{r}{4} \tilde{q}_1, \hspace{1em}
  \tilde{p}_2 = r^{- 4} p_2 + \frac{r^{- 1}}{2} p_1 \tilde{q}_1 + \frac{r}{4}
  \left( \tilde{q}_2 + \frac{r}{16} \tilde{q}_1^2 \right)
\end{equation}
with the mass $m$ in all arguments here.

\section{Small $z$ expansion\label{appMPT}}

By straightforward manipulations one can show that
\begin{equation}
  \langle \bar{\phi}^m \rangle = \frac{\langle \bar{\phi}^m P (
  \bar{\phi}) \rangle_{0, \bar{\phi}}}{\langle P ( \bar{\phi})
  \rangle_{0, \bar{\phi}}} \label{phimom}
\end{equation}
holds with a polynomial in $\bar{\phi}$ deriving from
\begin{equation}
  P ( \bar{\phi}) = P_0 ( \bar{\phi}) \langle \mathe^{- S_1}
  \rangle_{0, \eta}
\end{equation}
truncated at the desired order in $g_0^{1 / 2}$. The subscripts of the
averages refer to the parts of the action (\ref{S0phi}), (\ref{S0eta}) used.
To construct $P$ we perform Wick contractions with the $\eta$ propagator
\begin{equation}
  \tilde{\Delta}_0 (p) = \frac{1}{\hat{p}^2 + g_0^{1 / 2} (2 z_0^2 +
  \bar{\phi}^2) / (2 L^2)}, \hspace{1em} \Delta_0 (x) = \frac{1}{L^4}
  \sum_{p \neq 0} \mathe^{i p x} \tilde{\Delta}_0 (p),
\end{equation}
which will be expanded in $g_0^{1 / 2}$ in the end. The factor $P_0$ derives
from the Gaussian integral over $\mathe^{- S_{0, \eta}}$ and is given by
\begin{equation}
  \ln P_0 = \frac{1}{2} \sum_{k \geqslant 1} \frac{(- 1)^k}{k 2^k} g_0^{k / 2}
  (2 z_0^2 + \bar{\phi}^2)^k C_k (L) \label{lP0}
\end{equation}
with
\begin{equation}
  C_k = L^{- 2 k} \sum_{p \neq 0}  \frac{1}{( \hat{p}^2)^k} . \label{Ckdef}
\end{equation}
The dependence of $C_k$ on $L$ will be discussed in more detail below. By
computing connected graphs we obtain in addition
\begin{equation}
  \ln [P / P_0] =  - \frac{g_0}{8} X^2 + \frac{g_0^{3 / 2}}{12} 
  \bar{\phi}^2 L^2 \widetilde{\Delta_0^3} (0) +
  \frac{g_0^2}{48} \left[ L^4 \widetilde{\Delta_0^4} (0) + 3 X^2
  \widetilde{\Delta_0^2} (0) \right] + \Omicron (g_0^{5 / 2}) 
\end{equation}

with the same notation as in the (\ref{powtilde}) and the short hand
\begin{equation}
  X = L^2 \Delta_0 (0) = \sum_{k \geqslant 1} C_k \left[ - \frac{g_0^{1 /
  2}}{2} (2 z_0^2 + \bar{\phi}^2) \right]^{k - 1} .
\end{equation}
To compute the mass in a similar fashion we expand
\begin{equation}
  \langle \hat{p}^2_{\ast} \tilde{\Delta} (p_{\ast}) \mathe^{- S_1}
  \rangle_{0, \eta} = P ( \bar{\phi}) \times Q ( \bar{\phi})
\end{equation}
with $Q$ built from connected diagrams with two external lines at $p = \pm
p_{\ast}$,
\begin{eqnarray}
  Q & = & Y - \frac{g_0}{2 K_L} X Y^2 + \frac{g_0^{3 / 2}}{2 K_L} Y^2 
  \bar{\phi}^2  \widetilde{\Delta_0^2} (p_{\ast}) + \frac{g_0^2}{12 K_L}
  Y^2 \times \nonumber\\
  &  & \left[ 2 L^2 \widetilde{\Delta_0^3} (p_{\ast}) + 3 X
  \widetilde{\Delta_0^2} (0) + 3 X^2 Y / K_L \right] + \Omicron (g_0^{5/2})  
\end{eqnarray}
with
\begin{equation}
  Y = \hat{p}^2_{\ast} \tilde{\Delta}_0 (p_{\ast}) = \sum_{n \geqslant 0}
  \left[ - \frac{g_0^{1 / 2}}{2 K_L} (2 z_0^2 + \bar{\phi}^2) \right]^n
\end{equation}
evaluated up to the required order. We introduce additional constants for
\begin{equation}
  L^{2 k - 4} \widetilde{\Delta_0^k} (0) = D_k + g_0^{1 / 2} (2 z_0^2 +
  \bar{\phi}^2) D_k' + \Omicron (g_0), \hspace{1em} [\Rightarrow D_2
  \equiv C_2]
\end{equation}
and
\begin{equation}
  L^{2 k - 4} \widetilde{\Delta_0^k} (p_{\ast}) = D_k^{\ast} + g_0^{1 / 2} (2
  z_0^2 + \bar{\phi}^2) {D^{\ast}_k}' + \Omicron (g_0) .
\end{equation}
They are given by
\begin{equation}
  D_k = L^{- 2 k} \sum_{p_1 \ldots, p_k \neq 0} \delta_{\sum_i p_i, 0}
  \prod_{j = 1}^k \frac{1}{\hat{p}_j^2}
\end{equation}
and
\begin{equation}
  D_k' = - \frac{k}{2} L^{- 2 k - 2} \sum_{p_1 \ldots, p_k \neq 0}
  \delta_{\sum_i p_i, 0}  \frac{1}{( \hat{p}_1^2)^2} \prod_{j = 2}^k
  \frac{1}{\hat{p}_j^2}
\end{equation}
and corresponding formulas for $D_{k}^{\ast}$ and ${D_{k}^{\ast}}'$ with $p_{\ast}$
replacing zero for the total momentum. We now discuss the behavior of these
constants as far as they enter into our computation.

We first note that there are a number of universal logarithmic divergences
that enter into our final expansion coefficients $B_k$, like
\begin{equation}
  L \frac{\partial}{\partial L} C_2 = L \frac{\partial}{\partial L} D_2 =
  \frac{1}{8 \pi^2} + \Omicron (L^{- 2}) = L \frac{\partial}{\partial L}
  D_2^{\ast} . \label{C2log}
\end{equation}
The constants $D_3$ and $D_3^{\ast}$ are quadratically divergent, but their
difference obeys
\begin{equation}
  L \frac{\partial}{\partial L} (D_3 - D_3^{\ast}) = \frac{1}{64 \pi^2} +
  \Omicron (L^{- 2}) .
\end{equation}
Another combination that was found to occur in our final result is
\begin{equation}
  L \frac{\partial}{\partial L} \left( C_2^2 + \frac{4}{3} D_3' \right) = -
  \frac{1}{64 \pi^4} + \Omicron (L^{- 2}) .
\end{equation}
All other constants are non-universal (reflect the hypercubic lattice
and the discretization) and are
either divergent ($C_1, D_4, \text{$D_3 + D_3^{\ast}$}$) or finite. They all
drop out in our finite universal results for the $\beta$ function. The
behavior claimed above was checked by computing the constants numerically in
the same way as reported in the previous appendix. However, the universal
coefficients, which are the only feature needed here, should also be
computable more easily in the continuum.

We are now ready to compute
\begin{equation}
  \frac{z^2}{1 + z^2 / K_L} = g_0^{1 / 2} \frac{\langle P ( \bar{\phi}) Q
  ( \bar{\phi}) \bar{} \rangle_{0, \bar{\phi}}}{\langle P (
  \bar{\phi}) \bar{\phi}^2 \rangle_{0, \bar{\phi}}}
\end{equation}
and $g / z^4$ from (\ref{ging0}) and (\ref{phimom}). We have implemented all
the series in \tmtexttt{maple} [with independent codes in Berlin and Munich]
and obtain the corresponding truncated series
in $g_0^{1 / 2}$ with coefficients given above and in terms of the moments
\begin{equation}
  \mu_m = \langle \bar{\phi}^m \rangle_{0, \bar{\phi}}, \hspace{1em}
  \mu \equiv \mu_2 .
\end{equation}
Derivatives with respect to $z_0^2$ are taken with the help of
\begin{equation}
  \frac{\partial}{\partial z_0^2} \mu_m = - \frac{1}{2} (\mu_{m + 2} - \mu_m
  \mu_2) .
\end{equation}
Partial integration implies
\begin{equation}
  \mu_{m + 4} + 6 z_0^2 \mu_{m + 2} - 6 (m + 1) \mu_m = 0
\end{equation}
which allows us to express all moments by $\mu = \mu_2$ in the final result.
In this way we have arrived at the coefficients quoted under (\ref{B1res}),
(\ref{B2res}), (\ref{B3parts}) and (\ref{B3res}).

For $z_0^2 > 0$ the integral with (\ref{S0phi}) can be found under 3.323 in
{\cite{gradshteyn1980}} and yields the result
\begin{equation}
  \mu = 3 z_0^2 \left( \frac{K_{3 / 4} (3 z_0^4 / 4)}{K_{1 / 4} (3 z_0^4 / 4)}
  - 1 \right) \label{muexact}
\end{equation}
with the modified Bessel function $K_{\nu} (.)$ of index $\nu$. For moderately
negative $z_0^2$ we have simply summed the expansion for the integrals in
$z_0^2$ to sufficiently high order. To compare with ordinary perturbation
theory we need the perturbative expansion of $\mu$ for large $z_0^2$,
\begin{equation}
  \mu = \frac{1}{z_0^2} - \frac{1}{2 z_0^6} + \frac{2}{3 z_0^{10}} -
  \frac{11}{8 z_0^{14}} + \frac{34}{9 z_0^{18}} + \Omicron (z_0^{- 22}) .
  \label{muexp}
\end{equation}
\begin{figure}[htb]
\begin{center}
  \resizebox{0.8\textwidth}{!}{\includegraphics{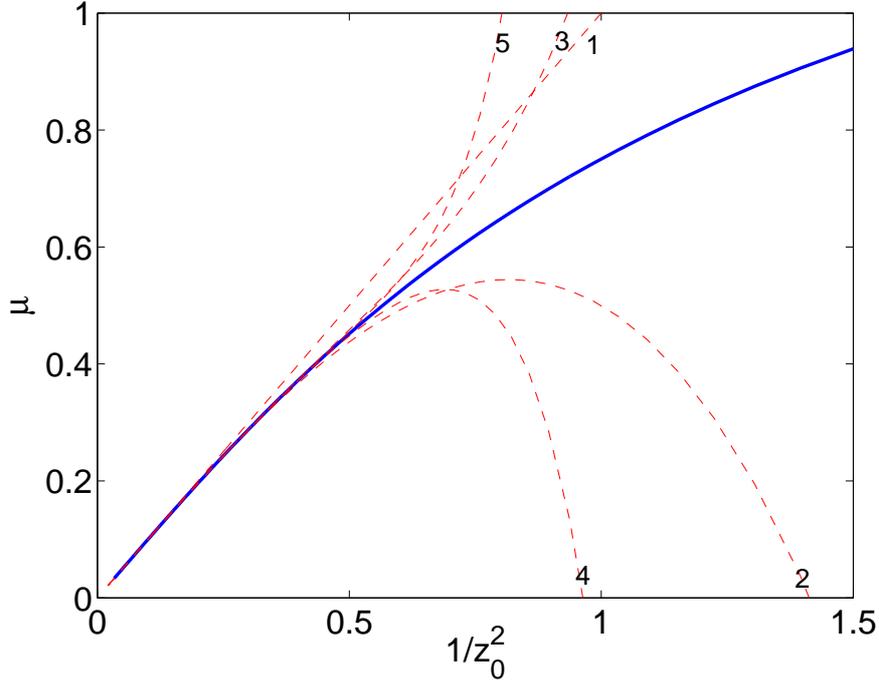}}
  \caption{The behavior of $\mu$ from (\ref{muexact}) and its perturbative
  approximations.\label{muplot}} 
\end{center}
\end{figure}

{\noindent}In fact this is a nice pedagogical example for the working of an
asymptotic series. In Fig. \ref{muplot} we see the exact $\mu$ of
(\ref{muexact}) together with various truncations of the series (\ref{muexp}).
There is a range around $z_0^{- 2} \gtrsim 1$ where the leading order alone is
the most decent approximation, a situation reminiscent of the behavior of the
perturbative series in Fig. \ref{rungz2}.


\begin{thebibliography}{10}
  \bibitem{Brezin:1976bp}E.~Brezin, J.~C. Le~Guillou, J.~Zinn-Justin, Field
  Theoretical Approach to Critical Phenomena in \tmtextit{Phase Transitions
  and Critical Phenomena}, Vol.6, London 1976, 125.
  
  \bibitem{Luscher:1987ay}M.~L\"uscher, P.~Weisz, Scaling Laws and
  Triviality Bounds in the Lattice phi**4 Theory. 1. One Component Model in
  the Symmetric Phase, Nucl. Phys. B290 (1987) 25.
  
  \bibitem{Wolff:2009ke}U.~Wolff, Precision check on triviality of $\phi^4$
  theory by a new simulation method, Phys. Rev. D79 (2009) 105002.
  
  \bibitem{Montvay:1988uh}I.~Montvay, G.~M\"unster, U.~Wolff, Percolation
  Cluster Algorithm and Scaling Behavior in the four-dimensional Ising Model,
  Nucl. Phys. B305 (1988) 143.
  
  \bibitem{Aizenman:1981du}M.~Aizenman, Proof of the Triviality of phi**4
  in D-Dimensions Field Theory and Some Mean Field Features of Ising Models
  for D$>$4, Phys. Rev. Lett. 47 (1981) 1.
  
  \bibitem{Lebowitz:1974}J.~L. Lebowitz, GHS and other inequalities,
  Commun. Math. Phys. 35 (1974) 87.
  
  \bibitem{Luscher:1991wu}M.~L\"uscher, P.~Weisz, U.~Wolff, A Numerical
  Method to compute the running Coupling in asymptotically free Theories,
  Nucl. Phys. B359 (1991) 221.
  
  \bibitem{Luscher:1992an}M.~L\"uscher, R.~Narayanan, P.~Weisz, U.~Wolff,
  The Schr\"odinger Functional: A renormalizable Probe for non-Abelian Gauge
  Theories, Nucl. Phys. B384 (1992) 168.
  
  \bibitem{Brezin:1985xx}E.~Brezin, J.~Zinn-Justin, Finite Size Effects in
  Phase Transitions, Nucl. Phys. B257 (1985) 867.
  
  \bibitem{Bode:1999sm}A.~Bode, P.~Weisz, U.~Wolff, Two loop Computation
  of the Schr\"odinger Functional in Lattice QCD, Nucl. Phys. B576 (2000) 517.
  
  \bibitem{gradshteyn1980}I.~S. Gradshteyn, I.~M. Ryzhik, Table of
  Integrals, Series, and Products, Academic Press, Boston, 1980.
\end{thebibliography}
\end{document}